\documentclass[reprint,eqsecnum,floats,aps,amsmath,amssymb,nofootinbib,prd,onecolumn, showpacs]{revtex4-2}
\usepackage{graphicx,physics}
\usepackage{amsmath,amssymb,mathtools,mathrsfs}
\usepackage{hyperref}
\usepackage{graphicx}
\usepackage{subfigure}
\usepackage{arydshln}
\usepackage{xcolor}
\usepackage{braket}
\usepackage{tensor}
\usepackage{enumitem,array,textcomp}
\usepackage[utf8x]{inputenc}
\usepackage{bigints}

\begin{document}
\title{Effects of the inflaton potential on the primordial power spectrum in Loop Quantum Cosmology scenarios}

\author{Beatriz Elizaga Navascu\'es}
\email{bnavascues@lsu.edu}
\affiliation{Department of Physics and Astronomy, Louisiana State University, Baton Rouge, LA 70803-4001, USA}

\author{Guillermo A. Mena Marug\'an}
\email{mena@iem.cfmac.csic.es}
\affiliation{Instituto de Estructura de la Materia, IEM-CSIC, Serrano 121, 28006 Madrid, Spain}

\author{Jes\'us Y\'ebana Carrilero}
 \email{jyacbana@ucm.es}
\affiliation{Instituto de Estructura de la Materia, IEM-CSIC, Serrano 121, 28006 Madrid, Spain}

\begin{abstract}
In scenarios of physical interest in Loop Quantum Cosmology, with a preinflationary epoch where the kinetic energy of the inflaton dominates, the analytic study of the dynamics of the primordial fluctuations has been carried out by neglecting the inflaton potential in those stages of the evolution. In this work we develop approximations to investigate the influence of the potential as the period of kinetic dominance gives way to the inflationary regime, treating the potential as a perturbation. Specifically, we study how the potential modifies the effective mass that dictates the dynamics of the scalar perturbations in the preinflationary epochs, within the framework of the so-called hybrid prescription for Loop Quantum Cosmology. Moreover, we motivate and model a transition period that connects the kinetically dominated regime with inflation, allowing us to study the interval of times where the contribution of the potential is no longer negligible but an inflationary description is not yet valid. Finally, we include the main modifications coming from a slow-roll correction to a purely de Sitter evolution of the perturbations during inflation. We analytically solve the dynamics of the perturbations in each of these different epochs of cosmological evolution, starting from initial conditions fixed by the criterion of asymptotic Hamiltonian diagonalization. This enables us to compute and quantitatively analyze the primordial power spectrum in a specific case, using a quadratic inflaton potential.
\end{abstract}

\maketitle

\section{\label{sec:level1}Introduction}

Over the last two decades, Loop Quantum Cosmology (LQC) \cite{LQC1,LQC2,LQC3} has settled as an active discipline for the consideration of quantum gravitational phenomena in the Early Universe \cite{Edreview,Hybridreview,Ivanreview}. This theoretical formalism has sensibly incorporated corrections from the theory of Loop Quantum Gravity \cite{LQG1,LQG2} to the evolution of the primordial fluctuations around the cosmological metric, as well as around the scalar field that conventionally drives a period of cosmic inflation. Part of these fluctuations, considered as cosmological perturbations, are widely believed to be the seeds for the formation of the large-scale structure of the Universe that we observe nowadays \cite{structures,mukhanov1,Cosmology}. In fact, their classical description --within General Relativity (GR)-- in early epochs using the standard cosmological model provides a primordial power spectrum that has been tested to a great degree of accuracy by high precision observations of the Cosmic Microwave Background (CMB) \cite{Planck,Planck-inf}. The best fit of the cosmological parameters of this model to the CMB observations however displays certain statistical anomalies in the low multipole region of the angular spectrum of temperature and polarization anisotropies \cite{Anomalies,CrisisWill,Crisis,AshtPRLLast,IvanAnomalies1}. Even though this region is strongly affected by cosmic variance, in principle it should not be ruled out that these anomalies may be an indicator of new physics beyond the standard model (and even beyond GR). In this context, LQC has shown to be a promising candidate to alleviate the observational tensions by invoking the quantum nature of gravity at the Planck scale \cite{Ashtekarlast,IvanFront}.

In contrast with previous attempts to ease the CMB anomalies within GR \cite{CPKL,CCL,DVS1,DVS2,Ramirez,PL,HHHL}, via the introduction of scalar field regimes with kinetic dominance in preinflationary periods, in LQC there exists a natural curvature scale of Planckian order that turns out to be important for the  perturbations \cite{JCAPGB}. More concretely, the (so-called effective) cosmological dynamics considered in LQC describes universes, minimally coupled to an inflaton field, with an evolution that connects in a smooth way contracting with expanding homogeneous and isotropic spacetimes \cite{Taveras}. This happens through a bouncing mechanism of quantum origin that replaces the classical singularity \cite{APS1,APS2,MMO}. This bounce is then shortly followed by a classical regime. In situations of phenomenological interest in view of their compatibility with observations, while allowing for falsifiable quantum corrections, the kinetic energy of the inflaton dominates in this region after the bounce, until the Universe enters a stage when the cosmological dynamics transitions into a sufficiently long period of slow-roll inflation \cite{IvanPhase,wang2,Louko}.The dynamics of the Fourier modes of the primordial fluctuations encode these modifications to the standard cosmological model via an effective time-dependent mass that incorporates LQC corrections \cite{massmax}. The relevance of these corrections for modes in the observable window of the CMB depends on the duration of the inflationary period \cite{AS,LB,BBMM,BGR,BLGR}. If this duration is not excesively large, the corrections can leave a measurable imprint, especially at large angular scales \cite{IvanPhase}. The effective mass of the scalar perturbations in LQC is the counterpart of the standard ratio $-\ddot{z}/z$ found in GR, where the dot denotes the derivative with respect to the conformal time and $z=a^2\dot{\phi}/\dot{a}$, with $a$ and $\phi$ denoting the scale factor and the scalar inflaton field, respectively. The changes between the dynamics of LQC and GR explain the modifications that appear in this mass, that can be drastic. Moreover, the evolution of the perturbations is especially sensitive to the behavior of this mass. In particular, at the LQC bounce it reaches (in absolute value) a universal maximum of Planckian order, as one could expect from quantum gravity considerations, and then rapidly decreases to very small, nearly classical values during the kinetic epoch. Furthermore, it displays a substantial change of behavior at an instant of time (that depends on the initial conditions) happening just before slow-roll inflation starts, epoch when the mass becomes increasingly large (in norm), as it is standard in conventional slow-roll scenarios. The value of the mass at that instant defines a curvature scale at the onset of inflation, which is also present in pure GR models with a period of kinetic dominance preceding slow-roll inflation \cite{CPKL,JCAPGB}. However, the Planckian scale at the curvature maximum is a genuine feature of LQC and its existence affects the primordial power spectrum, at least for reasonable choices of a vacuum state for the perturbations.

With a general understanding of the origin of the possible modifications to the primordial power spectrum, confronting the predictions from LQC with the CMB observations requires performing a full statistical analysis of the best fit of the cosmological parameters that appear in the theory. Recent efforts have been made in this direction \cite{Ashtekarlast,MJRMonteC} by two different strategies in LQC that lead to power suppression at low observable scales for the perturbations, as indicated by the anomalies: The dressed metric \cite{AAN,AAN1,AAN2} and hybrid approaches \cite{hyb-pert1,MSCastello,hybpred1}. However, these studies start from purely numerical expressions for the primordial power spectrum (or certain parameterizations of them) and, in part because of this, a statistical analysis that properly includes all of the free parameters that are relevant in LQC, including e.g. the value of the inflaton field at the bounce, is still missing\footnote{Some progress has been achieved for hybrid LQC in Ref. \cite{MJRMonteC} using certain parametrizations for the power spectrum and with a specific choice of vacuum state along the lines proposed in Ref. \cite{NO}. This vacuum is not especially responsive to the quantum bounce and differs from the vacuum that we will consider in this work.}. Such an analysis is crucial for a comprehensive understanding of the physical consequences of an LQC description of the Early Universe. 

A first approach to the analytic characterization of the primordial power spectrum in the case of hybrid LQC has been developed in Ref. \cite{JCAPGB}, starting with a physically well-motivated choice of vacuum state and leading to power suppression at low scales. These investigations have considerably simplified the dynamics of the perturbations by completely ignoring the effects of the inflaton potential in the earliest epochs of kinetic dominance, and then treating the slow-roll inflationary period as an exact de Sitter phase. The resulting analytic formula for the power spectrum, though interesting for qualitative analyses, is clearly not realistic enough for observational purposes. The aim of this work is to fill this gap in two fronts. On the one hand, we incorporate corrections to the dynamics of the perturbations taking into account that the inflaton potential becomes increasingly non-negligible in the epoch of kinetic dominance, as one approaches the inflationary regime. On the other hand, we apply the slow-roll approximation at first order to calculate the resulting power spectrum, reflecting the fact that the inflationary epoch is not exactly of de Sitter type. 

These improvements pave the road for a more detailed discussion of the LQC effects on the power spectrum and their comparison with observations, including a full statistical analysis, which we do not perform in this work but will be the subject of future research. Our results are intimately attached to the hybrid formalism for LQC and a specific criterion for the choice of vacuum of the cosmological perturbations, used in Ref. \cite{JCAPGB}. Nonetheless, extensions to other LQC formalisms and vacua are possible, for instance to the dressed metric formalism with the same criterion for the vacuum choice, as advocated in Ref. \cite{AMMVB}. Moreover, our study will focus on a quadratic inflaton potential, although a similar line of arguments can be followed for other potentials (in this context, let us recall that the effective mass of the perturbations was investigated in Ref. \cite{IMM} for dressed metric and hybrid LQC in the cases of the Starobinsky potential and some exponential potentials). It is worth commenting that a quadratic potential is not currently favored by a statistical analysis of the observations of the CMB within the standard inflationary model (see e.g. the results of the Planck mission in Ref. \cite{Planck-inf}). One should, however, be extremely careful when extrapolating this preference to the study of perturbations in LQC, a framework in which a similar statistical analysis of the data has not been completed, including different families of potentials and adapting the choice of vacuum state to the new situation, which modifies the standard scenario. In addition, it has been argued that genuine LQC modifications to the power spectrum should not vary much with the inflaton potential, at least qualitatively \cite{IvanPhase,Bon}. It therefore seems reasonable to study first the simplest case of a quadratic potential, and confront the extension of this study to other potentials in future investigations. 

To include the effects of an inflaton potential that is not fully negligible during kinetic dominance, we will employ the analysis developed in Ref. \cite{Rafael}. That reference studies the modifications produced by a generic inflaton potential in the quantum expectation values on the LQC geometry that determine the effective mass of the perturbations, the definition of the conformal time, and the initial conditions for the choice of vacuum state used in Ref. \cite{JCAPGB}. These studies provide, in particular, general formulas for these modifications in regimes where LQC is indistinguishable from GR. In the present work, we particularize this analysis to states peaked on effective trajectories of LQC \cite{LQC3,Taveras,hybpred1} in cases of phenomenological interest, working out all relevant formulas in regions of kinetic dominance with classical behavior of these effective trajectories, and completing all calculations for a quadratic potential. Moreover, we use and apply the previous results of Ref. \cite{Rafael} about the corrections from this potential to the vacuum state, with the aim of computing the associated primordial power spectrum, which is our primary goal here. 

The structure of this work is the following. In Sec. II, we  explicitly compute corrections to the effective time-dependent mass of the perturbations in the epoch of kinetic dominance, owing to the presence of the inflaton potential at first order. After checking the accuracy of this approximation, we motivate the consideration of a transition regime for the dynamics of the perturbations as they enter the inflationary epoch. In Sec. III we present our criterion for the choice of a vacuum state for the perturbations, and then we analytically solve their dynamical equations using the initial data determined by such choice. This allows us to find, in Sec. IV, an analytic formula for the primordial power spectrum associated with our choice of vacuum, which leads to power suppression. With this formula, we evaluate the spectrum explicitly and check how it changes if one ignores separately any of the types of corrections that we introduce in this work: Contributions of the potential during kinetic dominance, a transition epoch, and the slow-roll effects. Finally, in Sec. V we summarize our results and give an outlook for them. We work in Planck units, with $\hbar = G = c = 1$.

\section{The preinflationary effective mass}

Let us start by summarizing the effective dynamics of cosmological perturbations of scalar type in the hybrid approach to LQC. This formalism is based on a canonical quantization of the relativistic system formed by a spatially flat, homogeneous, and isotropic spacetime background minimally coupled to the inflaton field $\phi$, with inhomogeneous perturbations truncated at quadratic order in the action. In the hybrid approach, the background is quantized using LQC techniques, whereas the perturbations are quantized using a more standard, Fock representation. Treating the Mukhanov-Sasaki gauge invariant \cite{Mukhanov,Sasaki,Sasaki2} and its canonical momentum as fundamental fields for the description of the scalar perturbations, it is possible to Abelianize the constraints of the system and render the homogeneous mode of the Hamiltonian constraint as the only nontrivial one to impose. Its vanishing as a quantum operator on a wide class of physical states eventually leads to the following effective equations that dictate the dynamics of the Fourier modes $v_{\vec{k}}$ of the Mukhanov-Sasaki field:
\begin{equation}\label{effMS}
\ddot{v}_{\vec{k}} + \left( k^{2} + s \right)v_{\vec{k}} = 0,
\end{equation}
where $k$ is the Euclidean norm of the wavevector $\vec{k}$ (belonging to $\mathbb{R}^3\setminus \{0\}$) of the perturbation mode, the dot denotes derivative with respect to the conformal time $\eta$, as we mentioned in the Introduction, and $s$ is a time-dependent mass term that encodes the LQC modifications on the otherwise relativistic evolution. For additional details on the derivation of these equations, we refer the reader e.g. to Refs. \cite{Hybridreview,MSCastello}. Physically preferred choices of states for the homogeneous LQC geometry lead to an effective mass $s$ that has exactly the same functional dependence as in GR on the inflaton, the scale factor, and their canonical momenta \cite{massmax}. However, the evolution of these quantities in LQC differs from the relativistic one when the energy density of the inflaton approaches a few percentages of the Planck density, leading to the celebrated bounce. For the effective mass, this bounce produces a positive maximum, which is followed in the expanding branch by a rapid decrease that soon converges to a relativistic behavior, where the kinetic energy of the inflaton dominates. In particular, only after 0.4 cosmic Planck seconds from the bounce, the evolution of $s$ turns out to be indistinguishable from its relativistic counterpart \cite{JCAPGB}. Moreover, the contribution from the inflaton potential $W(\phi)$ at the bounce and shortly after it is completely negligible in the considered family of LQC geometries \cite{IvanPhase}. Nevertheless, the relativistic epoch of kinetic dominance that follows the bounce eventually ends and gives way to a period of slow-roll inflation. The purpose of this section is thus to analytically characterize the dominant contributions to the effective mass $s$ of the potential during the kinetic epoch, until the system experiences the transition to the inflationary regime, where $W(\phi)$ completely dominates the dynamics.

\subsection{Corrections to the mass at first order in the potential}

Let us review how to introduce first order corrections to the effective mass $s$ produced by the potential $W(\phi)$ in situations where the evolution of the background geometry with respect to $\phi$ is generated by a Hamiltonian $\mathcal{H}_0$ of the form
\begin{equation}\label{H0}
\mathcal{H}_0=\mathcal{H}_0^{(F)}-W(\phi)\frac{V^2}{\mathcal{H}_0^{(F)}}+\mathcal{O}(W^2),\qquad \mathcal{H}_0^{(F)}=\mathcal{H}_0^{(F)}(\pi_V, V),
\end{equation}
where $V$ is the physical volume (of a cubic cell with edges of coordinate length $2\pi$, namely $(2\pi a)^3$), $\pi_{V}$ is its canonical momentum, and the symbol $\mathcal{O}$ stands for terms of the order of its argument or smaller. We find this type of Hamiltonian at early times for the choice of LQC states that we consider in this work, peaked on trajectories that can be modelled by an effective Hamiltonian dynamics \cite{Taveras,hybpred1,qcCastello}. Furthermore, a function of the form \eqref{H0} also generates the evolution with respect to $\phi$ of the cosmological geometry in GR in regimes where $W(\phi)$ is negligible (a setting which reproduces very well the effective LQC dynamics shortly after the bounce). 

The procedure to incorporate in $s$ the corrections coming from the presence of $W(\phi)$ at first order has been developed in full detail in Ref. \cite{Rafael}, in the contexts of hybrid LQC and of GR. Here, we adapt the discussion and formulas of that work to scenarios where the cosmological evolution is generated by a Hamiltonian of the sort of $\mathcal{H}_0$. In particular, Ref. \cite{Rafael} provides a formula for the effective mass in conformal time of the form $s = s^{(F)} + s^{(W)} + \mathcal{O}(W^{2})$, where the first term refers to the effective mass in the ``free" scenario with vanishing inflaton potential, and the second one is the first order correction arising from the potential. Explicitly, we have
\begin{eqnarray}
&&s^{(F)}(\eta)=\frac{1}{3\pi}\frac{(\mathcal{H}_0^{(F)})^2}{V^{4/3}}\Bigg |_{\phi=\phi^{(F)}(\eta)},\\ \label{sW} && s^{(W)}(\eta)=\frac{2}{3\pi}V^{2/3}\left[\frac{3}{8\pi}W'' +15W+3\sqrt{\frac{3}{\pi}}\frac{\Lambda_0^{(F)}}{\mathcal{H}_0^{(F)}}W'\right]\Bigg |_{\phi=\phi^{(F)}(\eta)}-\left\{s^{(F)},J^{(W)}(\eta)\right\}\bigg|_{\phi=\phi^{(F)}(\eta)},
\end{eqnarray}
where $\phi^{(F)}(\eta)$ is the free dependence of the scalar field on the conformal time $\eta$, and all phase space functions to the left of the symbol $|_{\phi=\phi^{(F)}(\eta)}$ must be evaluated on the trajectories generated by the free Hamiltonian $\mathcal{H}^{(F)}_0$, with evolution parameter given by $\phi$. Besides, the prime denotes the derivative with respect to the inflaton, $\{.,.\}$ is the Poisson bracket, $\Lambda_{0}^{(F)}$ is the effective counterpart of an LQC operator that equals $-\mathrm{sign}(\pi_V)\mathcal{H}_{0}^{(F)}$ in GR, and $J^{(W)}(\eta)$ is a time and phase space dependent function defined as
\begin{equation}\label{JW}
J^{(W)}(\eta)=\mathcal{H}_{0}^{(F)}\phi^{(W)}(\eta)-\int_{\phi_0}^{\phi}d\tilde{\phi}\,K(\phi, \tilde{\phi})W(\tilde{\phi}),
\end{equation}
where $\phi_0=\phi^{(F)}(0)$ and
\begin{equation}\label{Kgen}
K(\phi,\tilde\phi)=\sum_{n=0}^{\infty}\frac{1}{n!}\frac{(\phi-\tilde\phi)^{n}}{\mathcal{H}_0^{(F)}}\left\{V^2,\mathcal{H}_{0}^{(F)}\right\}.
\end{equation}
It is not difficult to check that this last quantity corresponds to the evolution of $V^{2}/\mathcal{H}_0^{(F)}$ from $\tilde{\phi}$ to $\phi$ in the free case with $W(\phi)=0$. Finally, $\phi^{(W)}(\eta)$ is the first order correction (caused by the potential) to the evolution of the inflaton field in conformal time, and it is given by
\begin{equation}\label{phiW}
\phi^{(W)}(\eta)=-\eta^{(W)}\big(\phi^{(F)}(\eta)\big)\dot{\phi}^{(F)}(\eta),
\end{equation}
where $\eta^{(W)}(\phi)$ is the first order correction to the inverse of the inflaton field as a function of $\eta$,
\begin{equation}\label{etaWgen}
\eta^{(W)}(\phi)=\frac{3}{2}\bigintssss_{\phi_{0}}^{\phi} \frac{d\tilde{\phi}}{\left[\mathcal{H}_0^{(F)}\right]^2}\left[\frac{V^{8/3}}{\mathcal{H}_0^{(F)}}W(\tilde{\phi})+\int_{\phi_0}^{\tilde{\phi}}d\phi^{\star}W(\phi^{\star}) \bigg(V^{2/3} \big\{K(\tilde\phi,\phi^{\star}),\mathcal{H}_0^{(F)}\big\}+ \mathcal{H}_0^{(F)}\big\{V^{2/3},K(\tilde\phi,\phi^{\star})\big\}\bigg)\right].
\end{equation}
In this formula, all functions of the geometry must be evaluated on the trajectories, parameterized by $\tilde\phi$, generated by the free Hamiltonian $\mathcal{H}^{(F)}_0$.

\subsection{Approximating the effective mass in the kinetic epoch}

As we already mentioned, less than half a Planck second (in cosmic time) after the bounce, the effective mass $s$ in hybrid LQC is practically indistinguishable from its counterpart in classical GR. Moreover, in the typical backgrounds of interest, the contribution from the inflaton potential is completely negligible in the region near the bounce, and remains small in the kinetically dominated relativistic period after it. Therefore, in the following we approximate the effective mass in this kinetic epoch treating the inflaton potential as a small correction, keeping only first order contributions of this potential to the mass. We call $s^{(W)}$ these contributions, that modify the effective mass $s^{(F)}$ of the classical relativistic period ignoring the potential. In this respect, $\mathcal{H}_{0}^{(F)} = \sqrt{12\pi}|\pi_{V}|V$ is the Hamiltonian that generates the evolution of the geometry with respect to $\phi$ in the absence of a potential \cite{Rafael}. We notice that, up to a constant positive factor and possibly a sign, this Hamiltonian is just the generator of dilations in the physical volume. Evaluating Eq. \eqref{Kgen} in this case, we obtain the following result:
\begin{equation}
 K(\phi, \tilde{\phi}) = \frac{V^{2}}{\mathcal{H}_{0}^{(F)}}\exp\left[4\sqrt{3\pi}\, \mathrm{sign}(\pi_{V})(\phi-\tilde{\phi})\right],
\end{equation}
which, in turn, leads to the following simplification of Eq. \eqref{etaWgen}:
\begin{align}\label{etaWGR}
\eta^{(W)}(\phi)=\frac{3}{2}\bigintsss_{\phi_{0}}^{\phi} d\tilde{\phi}\frac{V^{8/3}}{\left[\mathcal{H}_0^{(F)}\right]^3}\left[W(\tilde{\phi})+\frac{8}{3}\sqrt{3\pi}\, \mathrm{sign}(\pi_{V})\int_{\phi_0}^{\tilde{\phi}}d\phi^{\star}W(\phi^{\star}) e^{4\sqrt{3\pi} \mathrm{sign}(\pi_{V})(\tilde\phi-\phi^{\star})}\right].
\end{align}
On the other hand, the relativistic dynamics of a flat, homogeneous, and isotropic spacetime minimally coupled to a massless scalar field is exactly solvable. It leads to the following evolution in conformal time (see e.g. Ref. \cite{JCAPGB}):
\begin{eqnarray}\label{Vpi}
&&V(\eta) = V_{0}\left(1+\frac{2\mathcal{H}_{0}^{(F)}}{\sqrt{3\pi}V_{0}^{2/3}}\eta\right)^{3/2},\qquad 
\pi_{V}(\eta) =- \frac{\mathcal{H}_{0}^{(F)}}{2\sqrt{3\pi}V_0}\left(1+\frac{2\mathcal{H}_{0}^{(F)}}{\sqrt{3\pi}V_{0}^{2/3}}\eta\right)^{-3/2},\\ \label{phiFeta}
&&\phi^{(F)}(\eta) - \phi_{0} = \sqrt{\frac{3}{16\pi}}\ln\left(1+\frac{2\mathcal{H}_{0}^{(F)}}{\sqrt{3\pi}V_{0}^{2/3}}\eta\right), 
\end{eqnarray}
where $V_0=V(0)$ and $\mathcal{H}_{0}^{(F)}$ is a constant of motion equal to the canonical momentum of the scalar field. If we introduce all of these expressions in Eq. \eqref{sW} (together with definitions \eqref{JW} and \eqref{phiW}), we obtain the following formula for the first order correction to the time-dependent mass $s$ in GR owing to the inflaton potential:
\begin{eqnarray}\label{sWGR}
s^{(W)}(\eta)&=&\frac{2}{\pi}\left(V_0^{2/3}+\frac{2\mathcal{H}_{0}^{(F)}}{\sqrt{3\pi}}\eta\right)\left[5W\big(\phi^{(F)}(\eta)\big)+\sqrt{\frac{3}{\pi}}W'\big(\phi^{(F)}(\eta)\big)+\frac{W''\big(\phi^{(F)}(\eta)\big)}{8\pi}\right]\nonumber \\ \nonumber &+&\frac{16}{3\pi}\sqrt{\frac{\pi}{3}}\left(V_0^{2/3}+\frac{2\mathcal{H}_{0}^{(F)}}{\sqrt{3\pi}}\eta\right)\int_{\phi_0}^{\phi^{(F)}(\eta)}d\tilde\phi \,W(\tilde\phi)e^{-4\sqrt{3\pi}\left[\phi^{(F)}(\eta)-\tilde\phi\right]}\\  &+&\frac{4}{3\pi}\sqrt{\frac{1}{3\pi}}\frac{\left[\mathcal{H}_{0}^{(F)}\right]^3}{V_0^2}\left(1+\frac{2\mathcal{H}_{0}^{(F)}}{\sqrt{3\pi}V_{0}^{2/3}}\eta\right)^{-3} \eta^{(W)}\big(\phi^{(F)}(\eta)\big),
\end{eqnarray}
where  $\eta^{(W)}$ is computed using Eqs. \eqref{Vpi} and \eqref{phiFeta} (replacing $\phi^{(F)}(\eta)$ with $\tilde\phi$ in the last of those equations) to write the functions of the geometry appearing in the integrand of Eq. \eqref{etaWGR} in terms of free trajectories parameterized by $\tilde{\phi}$.

The expression that we have obtained for $s^{(W)}$ is valid for any choice of inflaton potential. For concreteness, in the remainder of this work we will focus our discussion on a quadratic potential: $W(\phi)=m^2 \phi^2/2$. A direct computation then shows that
\begin{eqnarray}\label{sWGRm}
s^{(W)}(\eta)&=&\frac{2}{\pi}m^2\left(V_0^{2/3}+\frac{2\mathcal{H}_{0}^{(F)}}{\sqrt{3\pi}}\eta\right)\left[\frac{5}{2}\phi^{(F)}(\eta)^2+\sqrt{\frac{3}{\pi}}\phi^{(F)}(\eta)+\frac{1}{8\pi}\right]\nonumber \\ \nonumber & -& \frac{2m^2}{9\pi}\left(V_0^{2/3}+\frac{2\mathcal{H}_{0}^{(F)}}{\sqrt{3\pi}}\eta\right)\left[\left(1+\frac{2\mathcal{H}_{0}^{(F)}}{\sqrt{3\pi}V_{0}^{2/3}}\eta\right)^{-3}\bigg(\phi_0^2-\frac{\phi_0}{\sqrt{12\pi}}+\frac{1}{24\pi}\bigg)-\phi^{(F)}(\eta)^2+\frac{\phi^{(F)}(\eta)}{\sqrt{12\pi}}-\frac{1}{24\pi}\right]\\  &+&\frac{4}{3\pi}\sqrt{\frac{1}{3\pi}}\frac{\left[\mathcal{H}_{0}^{(F)}\right]^3}{V_0^2}\left(1+\frac{2\mathcal{H}_{0}^{(F)}}{\sqrt{3\pi}V_{0}^{2/3}}\eta\right)^{-3} \eta^{(W)}_m\big(\phi^{(F)}(\eta)\big),
\end{eqnarray}
for the quadratic potential, with
\begin{eqnarray}\label{etaWm}
\eta^{(W)}_{m}(\phi) &=& \frac{3m^{2}}{8\sqrt{3\pi}}\frac{V_{0}^{8/3}}{\left[\mathcal{H}_{0}^{(F)}\right]^{3}}\bigg[ e^{4\sqrt{3\pi}(\phi-\phi_{0})/3}\left( \phi_{0}^{2} - \frac{\phi_{0}}{2\sqrt{3\pi}} + \frac{1}{24\pi} \right) -\frac{9}{8}\left( \phi_{0}^{2} - \frac{3}{8\sqrt{3\pi}}\phi_{0} + \frac{3}{128\pi} \right) \nonumber \\ 
&+& \frac{1}{8} e^{16\sqrt{3\pi}(\phi-\phi_{0})/3}\left( \phi^{2} + \frac{5}{8\sqrt{3\pi}}\phi - \frac{47}{384\pi} \right) \bigg].
\end{eqnarray}

The formulas that we have obtained contain the main correction of the inflaton potential to the effective mass of the Mukhanov-Sasaki perturbations in the relativistic epoch of kinetic dominance. They can be used to explicitly compute an approximation of this time-dependent function as $s \approx s^{(F)}+s^{(W)}$ in this epoch. Here, the dominant mass term, for vanishing inflaton potential, is
\begin{equation}\label{sFGR}
s^{(F)}(\eta)=\frac{1}{4}\left(\eta+\frac{\sqrt{3\pi}}{{2\mathcal{H}_{0}^{(F)}}}V_{0}^{2/3}\right)^{-2}.
\end{equation}
In the rest of this section we study the range of applicability of this approximation, as well as possible improvements of it that allow to connect the kinetically dominated regime with slow-roll inflation.

\subsection{Validity of the approximation. Transition epoch}

In order to analyze the goodness of the approximation $s \approx  s^{(F)}+s^{(W)}$ in the classical epoch of kinetic dominance after the bounce, we are going to compare it with its exact counterpart in GR, given by \cite{massmax}:
\begin{eqnarray}
&&s_{\mathrm{GR}} = \frac{4}{3\pi}V^{2/3} \left[ \pi^2\frac{\dot{\phi}^2}{V_0^{2/3}} - W\left( \phi \right) \right] + \mathcal{U}, \\ \nonumber &&
\mathcal{U} = \frac{12}{\pi}V^{2/3} \left[ \frac{W^{\prime \prime}( \phi )}{48\pi} +  W( \phi ) + \frac{V \dot{\phi}}{\dot{V}}W^{\prime}( \phi ) - \frac{6V^{8/3}}{\pi\dot{V}^2}W^{2}( \phi ) \right].
\end{eqnarray}
The evolution of $V$ and $\phi$ is obtained by solving the Friedmann and Klein-Gordon equations, conveniently rewritten as
\begin{equation}\label{Fried}
\dot{V}=V^{4/3}\sqrt{\frac{6}{\pi}\left[2\pi^2 \frac{\dot{\phi}^2}{V^{2/3}}+W(\phi)\right]},\qquad \ddot{\phi}+\frac{2\dot{V}}{3V}\dot{\phi}+\frac{V^{2/3}}{4\pi^2}W'(\phi)=0.
\end{equation}
The solutions to these equations are completely fixed by the initial data $V_0$, $\phi_0$, and $\dot{\phi}(0)$. Here, we are only interested in solutions which connect with the effective LQC trajectories of phenomenological interest that we are considering. These experience bounces where the contribution of the potential to the energy density of the inflaton is completely negligible, and the dynamics there can be analytically solved \cite{wang2,JCAPGB}. Recalling that $0.4$ Planck seconds (in cosmic time) after the bounce are enough to neglect any LQC modification to the relativistic dynamics, we can use that instant to set the associated origin of conformal time, and fix $V_0$ and $\dot{\phi}(0)$ by requiring continuity with the corresponding effective LQC trajectory (computed by letting $W(\phi)$ vanish). Choosing e.g. positive $\dot{\phi}(0)$ and the scale factor to be equal to one at the bounce, we then get \cite{JCAPGB}
\begin{equation}\label{matchV}
V_0=8\pi^{3}\sqrt{1+\frac{96\pi}{25}\rho_c},\qquad \dot{\phi}(0)=\sqrt{ \frac{32\pi^4}{ V_0^{4/3}}\rho_c -\frac{V_0^{2/3}}{2\pi^2}W(\phi_0)},
\end{equation}
where $\rho_c  \approx  0.41$ is the value of the energy density at the bounce \cite{LQC3}. The only piece of data remaining to solve the cosmological dynamics in GR is $\phi_0$, namely the value of the inflaton field at $0.4$ Planck seconds after the bounce, in cosmic time. Besides, we need to fix the mass parameter $m$ that characterizes our quadratic choice for the inflaton potential. For concreteness, we choose the values $\phi_0=1.22$ (for which the corresponding value of the inflaton at the bounce is $0.97$) and $m = 1.2 \times 10^{-6}$. It is well known that choices similar to this one lead to a kinetically dominated bounce, together with an inflationary period that contains just enough e-folds to be compatible with the CMB observations while leaving room for LQC modifications at low observable scales $k$ \cite{IvanPhase,hybpred2}. Moreover, the above choice of values for the inflaton at the bounce and constant $m$ is precisely the choice made in the analytic and numerical studies of Refs. \cite{hybpred2,JCAPGB}, so that the results of those works can be used for comparison with the present analysis.

\begin{figure}[t]
    \includegraphics[scale = 0.60]{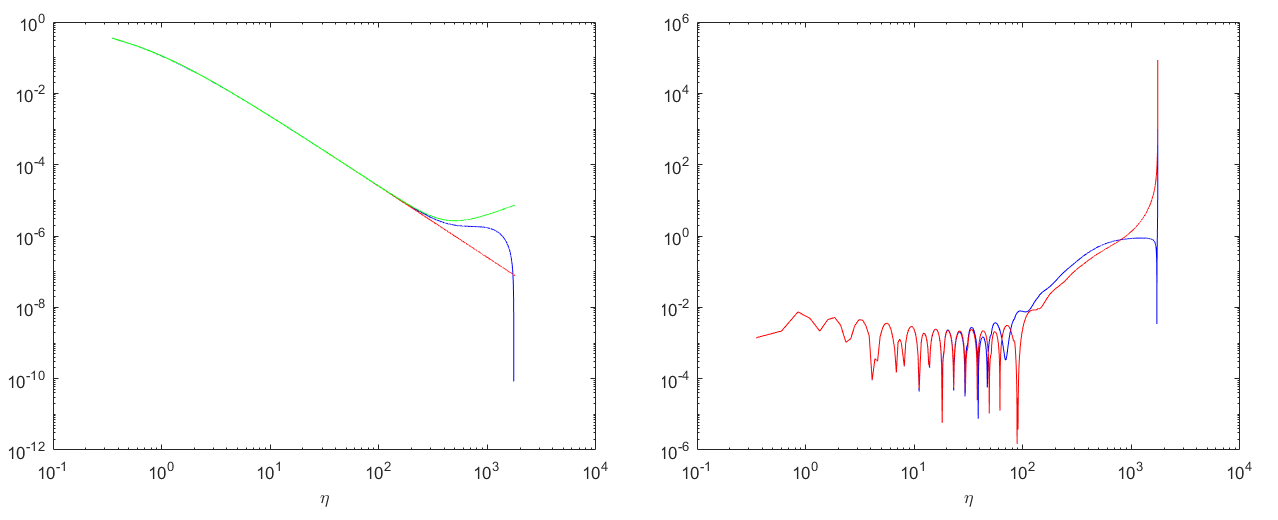}
    \caption{\textit{Left}: Numerical comparison between the curves $s_{\mathrm{GR}}(\eta)$ (blue), $s^{(F)}(\eta)$ (red), and $s^{(F)}(\eta)+s^{(W)}(\eta)$ (green). \textit{Right}: Relative error made when approximating the exact mass function $s_{\mathrm{GR}}(\eta)$ by $s^{(F)}(\eta)$ (blue) and by $s^{(F)}(\eta)+s^{(W)}(\eta)$ (red), in conformal time. In both panels, the axes are in a logarithmic scale.}
    \label{fig1}
\end{figure}

Starting with the set of data $\phi_0$ and $m$, we numerically integrate the relativistic dynamics of Eq. \eqref{Fried} using a Runge-Kutta method of 4th order. The resulting curve for the mass $s_{\mathrm{GR}}(\eta)$ is shown in blue color in the left panel of Fig. \ref{fig1}. In order to compare it with our approximation of the effective mass in the classical regime of kinetic dominance, namely $s  \approx  s^{(F)}+s^{(W)}$, it only makes sense to choose the same values of $\phi_0$ and $m$ (together with $V_0$ as in Eq. \eqref{matchV}) in order to evaluate Eqs. \eqref{sWGRm} and \eqref{etaWm}. Moreover, one can check that the matching between free solutions in GR and the considered effective LQC ones around the bounce leads to the following value of the constant of motion $\mathcal{H}_0^{(F)}$ \cite{JCAPGB}:
\begin{equation}\label{matchH}
\mathcal{H}_0^{(F)}=8\pi^3 \sqrt{2\rho_c}.
\end{equation}
In the left panel of Fig. \ref{fig1} we plot the resulting curves for $s^{(F)}$ and $s^{(F)}+s^{(W)}$ in red and green, respectively. They are both practically indistinguishable from their exact counterpart $s_{\mathrm{GR}}$ at early times, as we expect from the type of LQC scenarios that we are considering, since the potential contribution becomes completely negligible as one approaches the bouncing epoch. However, this contribution grows as one evolves to the future in the regime of kinetic dominance, until one has $s^{(F)}=s^{(W)}$ at a time around $\eta=430$. From this moment on, the first order correction owing to the presence of the potential becomes significantly relevant, modifying the behavior of the mass with respect to the free case. In this way, we obtain an approximate description for $s$ that fits its exact behavior for later times better than if we had completely ignored $W(\phi)$, as it was done in Ref. \cite{JCAPGB}.

The improvement in the approximation to the effective mass achieved by considering the first order correction $s^{(W)}$ is shown quantitatively in the right panel of Fig. \ref{fig1}. This panel displays the relative errors (with respect to the average of the two compared quantities) commited when approximating the exact relativistic mass $s_{\mathrm{GR}}$ by $s^{(F)}$ and $s^{(F)}+s^{(W)}$. We clearly see that including the contribution from $s^{(W)}$ keeps the approximation good until later times\footnote{The instant $\eta \approx 1800$ where the relative error of $s^{(F)}$ vanishes corresponds to a casual intersection between this curve and $s_{\mathrm{GR}}$ when inflation has already started, and has no physical significance.}. In particular, for times earlier than $\eta = 800$ (and later than the matching at $\eta=0$ with the bouncing epoch), the error made when approximating the effective mass by $s^{(F)}+s^{(W)}$ is less than $50\%$ (even if the mass is approaching a period with very small values). Beyond this instant, we see in the left panel of Fig. \ref{fig1} that the exact relativistic mass enters a plateau where it remains approximately constant until around $\eta=1300$. This moment can be understood as the end of kinetic dominance and the beginning of inflation, since one can check that the kinetic contribution to the energy density of the inflaton becomes comparable to its potential there. From this perspective, we can complete our analytic approximation to the effective mass, from $\eta=0$ (i.e. when LQC effects become negligible after the bounce) until $\eta=1300$, by matching the dynamics, at the moment when the approximation $s  \approx  s^{(F)}+s^{(W)}$ begins to fail, with a transition epoch between kinetic dominance and inflation where the mass is simply a constant, given by the value of $s_{\mathrm{GR}}$.

In summary, we have found that the following function approximates reasonably well the numerical behavior of the effective mass $s$ for the Mukhanov Sasaki perturbations in hybrid LQC, for times $\eta\in[0,1300]$:
\begin{equation}\label{prein}
s_{\mathrm{pre}} (\eta)= \bigg\lbrace \begin{array}{lcc}
s^{(F)}(\eta) + s^{(W)}(\eta), & \quad &  \eta \in \left[0, \eta_{t}\right],  \\
s_{c}=1.83\times 10^{-6}, & \quad & \eta \in \left[\eta_{t}, \eta_{i}\right] ,
\end{array}
\end{equation}
where $\eta_t=800$, $\eta_i=1300$, and we have taken the numerical value of $s_{\mathrm{GR}}(\eta_t)$ as the value for the mass in the transition epoch. For times earlier than $\eta=0$, the LQC effects are important in the cosmological dynamics. In this epoch near the bounce, we use a P\"oschl-Teller potential to approximate the effective mass. This has been shown to be a satisfactory approximation in that epoch (see e.g. Ref. \cite{JCAPGB}). On the other hand, we recall that for times later than $\eta_i$, inflation starts and eventually leads to a slow-roll regime. For simplicity, for $\eta>\eta_i$ we model the effective mass by its standard relativistic counterpart computed to first order in the slow-roll approximation. We expect that the errors made by adopting an instantaneous transition to the slow-roll inflationary regime and using standard slow-roll formulas in that epoch will not be especially important, considering that such transition is characterized by completely negligible values of $s$ (together with a change of sign in the effective mass), for which a variation affects mostly the evolution of modes with $k\ll 10^{-3}$, wavenumber scales which are too small to be observable in any plausible inflationary scenario.

\section{Evolution of the perturbations}

We have developed a (semi-)analytic description of the effective mass appearing in the dynamical equations \eqref{effMS} of the Mukhanov-Sasaki modes, which incorporates the first order effects of the inflaton potential in their evolution. The next natural step is to solve these dynamical equations, once a particular choice of initial conditions (or vacuum state) is provided for them. The criteria on which this choice is based must remain meaningful in the context of LQC. It has been widely debated that making this choice has important consequences in the shape of the resulting power spectrum \cite{hybpred2,NO,AG1,AG2,Mercelast,Mercelast2}. For example, one can ignore any physically relevant preinflationary effect by simply working with the standard Bunch-Davies vacuum \cite{BunchDavies} at the slow-roll inflationary epoch, or completely hide from sight the LQC phenomena at the bounce by choosing initial conditions that are natural if inflation is preceeded by just a classical epoch of kinetic dominance within GR \cite{CPKL,JCAPGB}. Bearing in mind the extensive (though yet inconclusive) literature on this topic, in this work we focus on a specific choice of vacuum state that is motivated by the criterion of asymptotic Hamiltonian diagonalization (AHD) \cite{BeaDiagonal}, specially well suited for hybrid quantization formalisms and, hence, for hybrid LQC \cite{Hybridreview,BeaDiagonal,BeaDiagonal2}. The origin of this criterion lies at the very quantization procedure for the Hamiltonian operator of the perturbations, and it naturally leads to a set of positive frequencies for them. Moreover, in effective scenarios, it selects solutions to the dynamical equations \eqref{effMS} with norms that display the minimal oscillations allowed in the evolution for asymptotically large $k$ \cite{BeaDiagonal2}, and certain non-oscillating (NO) behavior even beyond that asymptotic region. This feature, when valid for all modes that could be observable today, has been argued to be desirable for choices of a vacuum state in LQC that does not lead to power amplification. Because of these properties, we will call the vacuum selected by this criterion the NO-AHD vacuum.

This vacuum has already been considered in analytic studies that, within LQC, have proven an associated power supression starting from a scale cutoff of the order of the spacetime curvature at the bounce. To the best of our knowledge, this result is only shared (at least qualitatively) by the proposal of vacuum state for the perturbations put forward by Ashtekar and Gupt, in the so-called dressed metric prescription of LQC \cite{AG1,AG2}. We recall that, in this prescription, the perturbation modes have similar equations of motion as in hybrid LQC, but the effective mass differs in the two cases. For the dressed metric prescription, the effective mass has the same functional dependence on the background metric as in GR, but evaluated on the effective geometries that we are considering in LQC \cite{AAN,AAN1,AAN2,massmax}.

More specifically, the NO-AHD vacuum consists in (normalized) positive-frequency solutions to Eq. \eqref{effMS} of the form
\begin{equation}\label{muk}
\mu_{k} = \sqrt{-\frac{1}{2 \mathrm{Im}(h_{k})}}e^{i \int_{0}^{\eta}d\bar{\eta}\, \mathrm{Im}(h_{k})(\bar{\eta})},
\end{equation}
where the time-dependent function $h_{k}$ is a solution of the Riccati equation
\begin{eqnarray}\label{Riccati}
\dot{h}_{k} = k^{2} + s + h_{k}^{2}
\end{eqnarray}
that displays the following asymptotic expansion in the regime of large $k$ \cite{BeaDiagonal}:
\begin{equation}\label{asymptotich} 
\frac{1}{h_k}\simeq  \frac{i}{k}\left[1-\frac{1}{2k^2}\sum_{n=0}^{\infty}\left(\frac{-i}{2k}\right)^{n}\gamma_n \right].
\end{equation}
Here, $\gamma_{0}=s$ and the coefficients $\gamma_{n+1}$ for $n\geq 0$ are fixed by the recurrence relation
\begin{equation}
\gamma_{n+1} = -\dot{\gamma}_{n}+4s \left[\gamma_{n-1}+\sum_{m=0}^{n-3}\gamma_m \gamma_{n-(m+3)}\right]-\sum_{m=0}^{n-1}\gamma_m \gamma_{n-(m+1)}.
\end{equation}
These formulas lead to the standard choices of vacuum state in situations that are well understood in quantum field theory, such as the Poincar\'e and Bunch-Davies states in, respectively, Minkowski and de Sitter spacetimes \cite{BeaDiagonal}. They have also been shown to fix a well-behaved state in preinflationary epochs described by GR with a minimally coupled massless scalar field in kinetically dominated regimes \cite{JCAPGB}. In this work, we are interested in their application to the effective LQC epoch near the bounce. With an adequate choice of P\"oschl-Teller potential to approximate there the behavior of $s$, it was shown in Ref. \cite{JCAPGB} that the NO-AHD vacuum corresponds to positive-frequency solutions associated with the following choice of $h_k$:
\begin{equation}\label{hPTAHD}
h_k=-ik-2\alpha^2 x(1-x)	\frac{c\,d}{\alpha+ik}\frac{{}_2 F_1\left(c+1,d+1;2+ik/\alpha;x\right)}{{}_2 F_1\left(c,d;1+ik/\alpha;x\right)}, \qquad {\rm with} \qquad x=\left[1+e^{-2\alpha(\eta-\eta_B)}\right]^{-1}.
\end{equation}
Here, ${}_2F_1$ is the Gauss hypergeometric function, $\eta_B$ is the instant of the bounce in conformal time (recall that the origin of $\eta$ has been placed some tenths $t_0$ of a Planck second in cosmic time after the bounce), namely \cite{JCAPGB}
\begin{equation}
-\eta_B=\,{}_2 F_1\left(\frac{1}{6},\frac{1}{2};\frac{3}{2};-24\pi\rho_c t_0^2 \right)t_0,
\end{equation}
where in the case under study $t_0=0.4$, and we have defined
\begin{equation}\label{cd}
c=\frac{1}{2}\left(1+\sqrt{1+\frac{32\pi\rho_c}{3\alpha^2}}\right),\qquad d=\frac{1}{2}\left(1-\sqrt{1+\frac{32\pi\rho_c}{3\alpha^2}}\right), \qquad \alpha=-\frac{1}{\eta_B}\text{arcosh}\left(\frac{V_0^{2/3}}{4\pi^2}\right).
\end{equation}
The above solutions are valid from $\eta=\eta_B$ until $\eta=0$. At that end of the bounce epoch, they provide the initial data needed to integrate Eq. \eqref{effMS} in the relativistic preinflationary epoch, using our approximation \eqref{prein} for the effective mass $s$. \\

\subsection{Mode solutions in the kinetic epoch}

Let us study the solutions of the Mukhanov-Sasaki equations \eqref{effMS} when we approximate the effective mass as $s  \approx  s^{(F)} + s^{(W)}$, in the interval of conformal time $[0,\eta_t]$ of our model. For consistency, we want to obtain the expression of the solutions at first order in the inflaton potential. The general form of the positive-frequency solutions in the massless case (with $W(\phi)=0$) is exactly known in terms of Hankel functions of zeroth order, $H_0^{(1)}$ and $H_0^{(2)}$ \cite{CPKL,JCAPGB}. We call $\mu_k^{(F)}$ these free solutions. In general, they are completely fixed via Eq. \eqref{muk} by functions $h_k^{(F)}$ that solve the Riccati equation \eqref{Riccati} for $s=s^{(F)}$. In Ref. \cite{Rafael} one can find a detailed study of how the presence of the correction $s^{(W)}$ to the mass modifies the expression of these positive-frequency solutions to first order in the contributions of the inflaton potential. The modified solutions are obtained after inserting in Eq. \eqref{muk} the function $h_{k} = h_{k}^{(F)} + h_{k}^{(W)}$, where $h_{k}^{(W)}$ is given by
\begin{equation}\label{hW}
h_{k}^{(W)}(\eta) = \left[C_k^{(W)}+ \int_{0}^{\eta} d\bar{\eta}\, s^{(W)}(\bar{\eta})e^{-I_{h_{k}}^{(F)}(\bar{\eta})} \right] e^{I_{h_{k}}^{(F)}(\eta)},\qquad I_{h_{k}}^{(F)}(\eta) = 2\int_{0}^{\eta}d\bar{\eta}\,h_{k}^{(F)}(\bar{\eta}),
\end{equation}
and $C_k^{(W)}$ is an integration constant that we fix to zero in this work, so that $h_k$ coincides with the free solution $h_{k}^{(F)}$ to Eq. \eqref{Riccati} at the initial time $\eta=0$, when the contribution from the inflaton potential is completely negligible.

If we substitute the function $h_{k} = h_{k}^{(F)} + h_{k}^{(W)}$ in Eq. \eqref{muk} and expand to first order in the potential, we obtain the following formula for the positive-frequency solutions $\mu_k^{\mathrm{kin}}$ in the kinetically dominated epoch:
\begin{equation}
 \mu_{k}^{\mathrm{kin}}(\eta) = \mu_{k}^{(F)}(\eta)\left[ 1 + \mathfrak{F}_{k}^{(W)}(\eta) \right] ,\qquad \mathfrak{F}_{k}^{(W)}(\eta)=i\int_{0}^{\eta}d\bar{\eta}\, \textrm{Im}h_{k}^{(W)}(\bar{\eta})  - \frac{\textrm{Im}h_{k}^{(W)}(\eta)}{2\textrm{Im}h_{k}^{(F)}(\eta)},
\end{equation}
where $h_k^{(W)}$ is given by Eq. \eqref{hW} with $C_k^{(W)}=0$ and we recall that $\mu_{k}^{(F)}$ are the positive-frequency solutions to the Mukhanov-Sasaki equations with a mass term equal to the free relativistic function $s^{(F)}$ \cite{JCAPGB}. Explicitly, we obtain
\begin{eqnarray}\label{mukin}
\mu_{k}^{\mathrm{kin}}= \sqrt{\frac{\pi y}{4}} \bigg\lbrace C_{k}H_{0}^{(1)}(ky)\left[1+\mathfrak{F}_{k}^{(W)}\right]  
+ D_{k}H_{0}^{(2)}(ky)\left[1+\mathfrak{F}_{k}^{(W)}\right] \bigg\rbrace, \qquad y=\eta+ \sqrt{\frac{3}{8\pi\rho_c}}\frac{V_{0}^{2/3}}{8\pi^2},
\end{eqnarray}
where $C_k$ and $D_k$ are integration constants that satisfy $|D_{k}|^{2} - |C_{k}|^{2} = 1$, so that the solutions are normalized with respect to the Klein-Gordon product (up to higher than linear order in $W$). The choice of these constants for all modes is tantamount to the choice of a vacuum state for the perturbations. In this respect, we follow the NO-AHD proposal explained in the beginning of this section and impose as initial data the values at $\eta=0$ of the solutions $\mu_k$ and of their first time derivative obtained by inserting Eq. \eqref{hPTAHD} in Eq. \eqref{muk}. Taking into account that $\mathfrak{F}_{k}^{(W)}(0)=0$, this leads to 
\begin{eqnarray}\label{CD}
&&C_{k} =  \frac{1}{H_{0}^{(1)}(k/k_0)} \left[2\sqrt{\frac{k_0}{\pi}}\mu_{k}(0) 
- D_{k} H_{0}^{(2)}(k/k_0)\right],  \nonumber \\
&&D_{k} = \frac{i}{2}\sqrt{\frac{\pi}{k_0}}\left[ k H_{1}^{(1)}(k/k_0) \mu_{k}(0) 
- \frac{k_0}{2}H_{0}^{(1)}(k/k_0)  \mu_{k}(0)
+ H_{0}^{(1)}(k/k_0) \dot{\mu}_{k}(0) \right],
\end{eqnarray}
where we have defined the scale $k_0=8\pi^2 V_0^{-2/3}\sqrt{8\pi\rho_c/3}$. The resulting positive-frequency solutions $\mu_k^{\mathrm{kin}}$ are valid from $\eta=0$ to $\eta=\eta_t$ when, according to our approximation \eqref{prein}, the epoch that marks the transition to inflation starts. The behavior at $\eta=\eta_t$ of such mode solutions serves as initial data for the next period in the evolution of the perturbations.

\subsection{Mode solutions during the transition to inflation}

The epoch that connects the period of kinetic dominance with the inflationary regime corresponds to the interval of conformal time $[\eta_t,\eta_i]$. There, we have seen that the effective mass for the perturbations behaves approximately as a constant, $s_{c}= 1,83\times10^{-6}$. The Mukhanov-Sasaki equations \eqref{effMS} are straightforward to solve in this situation, yielding positive-frequency solutions $\mu_{k}^{\mathrm{c}}$ of the form
\begin{equation}\label{muc}
\mu_{k}^{\mathrm{c}}= \frac{iF_{k}}{\sqrt{2\kappa(k)}} e^{-i\kappa(k)(\eta-\eta_{t})} + \frac{iG_{k}}{\sqrt{2\kappa(k)}} e^{i\kappa(k)(\eta-\eta_{t})},\qquad \kappa(k)=\sqrt{k^2+s_c},
\end{equation}
where $F_k$ and $G_k$ are integration constants that satisfy $|F_k|^2 -|G_k|^2$, so that, once again, these solutions are normalized. Their specific value for our choice of vacuum state is computed imposing as initial data for the solutions the value at $\eta=\eta_t$ of the functions $\mu_k^{\mathrm{kin}}$ and of their time derivatives, determined by Eqs. \eqref{mukin} and \eqref{CD} in the preceeding cosmological epoch. This leads to the following constants:
\begin{equation}\label{FG}
F_{k} = -\sqrt{\frac{\kappa(k)}{2}} \left[ i\mu_{k}^{\mathrm{kin}}(\eta_{t}) - \frac{\dot{\mu}_{k}^{\mathrm{kin}}(\eta_{t})}{\kappa(k)} \right],
\qquad G_{k} =-\sqrt{\frac{\kappa(k)}{2}} \left[ i\mu_{k}^{\mathrm{kin}}(\eta_{t}) + \frac{\dot{\mu}_{k}^{\mathrm{kin}}(\eta_{t})}{\kappa(k)} \right].
\end{equation}
The resulting positive-frequency solutions describe the evolution of the perturbations from $\eta=\eta_t$ to $\eta=\eta_i$.

\subsection{Mode solutions in slow-roll inflation}

According to our approximations, we consider that the inflationary period starts at $\eta=\eta_i$, experiencing an expansion of quasi de Sitter type. The Universe then enters the celebrated period of slow-roll inflation, and by definition it holds as long as the following parameters are much smaller than one:
\begin{equation}
\epsilon_{W}=\frac{1}{16\pi}\left[\frac{W'(\phi)}{W(\phi)}\right]^2,\qquad \delta_{W}=\frac{1}{8\pi}\frac{W''(\phi)}{W(\phi)}.
\end{equation}
The slow-roll approximation at first order consists in treating these parameters as small constants, ignoring any non-linear contribution from them in the cosmological dynamics. Within this approximation, one can check that \cite{Baumann}
\begin{equation}\label{slowroll}
s_{\mathrm{GR}}  \approx  \left(\frac{\dot{a}}{a}\right)^2\left(-2+3\delta_W -5\epsilon_W\right),\qquad \frac{d}{d\eta}\left(\frac{a}{\dot{a}}\right)  \approx  \epsilon_W -1.
\end{equation}
It then follows that we can approximate the effective mass in the Mukhanov-Sasaki equations \eqref{effMS} in the slow-roll regime by the function
\begin{equation}
s_{\mathrm{sr}}=-\frac{1}{\left(\eta-\bar{\eta}_e\right)^{2}}\left(\nu^{2}-\frac{1}{4}\right), \qquad \nu=\sqrt{\frac{9}{4}+9\epsilon_W-3\delta_W},
\end{equation}
where $\bar{\eta}_e$ is a constant that can be identified, at any instant $\eta_e$ in the slow-roll interval, as
\begin{equation}\label{baretae}
\bar{\eta}_e=\eta_e +\frac{1}{1-\epsilon_W} \left.\left(\frac{a}{\dot{a}}\right)\right|_{\eta_e}.
\end{equation} 
We will take $\eta_e$ at the end of the slow-roll regime, defined as the instant when the slow-roll parameters stop behaving as constants, while still being much smaller than one. At that instant, we notice that $a/\dot{a}$ turns out to be negligible compared to $\epsilon^2_W$ and $\delta^2_W$ for typical inflationary scenarios (recall that the dot denotes the derivative with respect to the conformal time). Considering that we are working at first order in the slow-roll approximation, in the following analysis we work with the identification $\bar{\eta}_e=\eta_e$. With the above mass function, Eq. \eqref{effMS} can be written as a Bessel equation of order $\nu$ in the variable  $k(\eta_e-\eta)$. Therefore, positive-frequency solutions $\mu_k^{\mathrm{sr}}$ within the first order slow-roll approximation adopt the form
\begin{equation}\label{musr}
\mu_k^{\mathrm{sr}}=A_k\sqrt{\frac{\pi}{4}(\eta_e-\eta)}H_{\nu}^{(1)}\left[k(\eta_e-\eta)\right]+B_k\sqrt{\frac{\pi}{4}(\eta_e-\eta)}H_{\nu}^{(2)}\left[k(\eta_e-\eta)\right],
\end{equation}
where $H_{\nu}^{(1)}$ and $H_{\nu}^{(2)}$ are the Hankel functions of order $\nu $ and of the first and second kind, respectively, whereas $A_k$ and $B_k$ are integration constants. Again, normalization of the solutions demands that $|A_k|^2-|B_k|^2=1$. In our model, the initial data, given by the values at $\eta=\eta_i$ of the solutions $\mu_k^{\mathrm{c}}$ (and of their derivatives) specificied by Eqs. \eqref{muc} and \eqref{FG}, lead to the following expressions for these constants:
\begin{eqnarray}
A_{k} = &&-i\sqrt{\frac{\pi}{16}(\eta_e-\eta_{i})} \Bigg\lbrace kH_{\nu+1}^{(2)}\left[k(\eta_e-\eta_{i})\right] 
- kH_{\nu-1}^{(2)}\left[k(\eta_e-\eta_{i})\right] - \frac{H_{\nu}^{(2)}\left[k(\eta_e-\eta_{i})\right]}{(\eta_e-\eta_{i})} \Bigg\rbrace \mu_{k}^{\mathrm{c}}(\eta_{i}) \nonumber \\
&&
+i\sqrt{\frac{\pi}{4}(\eta_e-\eta_{i})}H_{\nu}^{(2)}\left[k(\eta_e-\eta_{i})\right] \dot{\mu}_{k}^{\mathrm{c}}(\eta_{i}),
\end{eqnarray}
\begin{eqnarray}\label{B}
B_{k} = &&i\sqrt{\frac{\pi}{16}(\eta_e-\eta_{i})} \Bigg\lbrace kH_{\nu+1}^{(1)}\left[k(\eta_e-\eta_{i})\right] 
- kH_{\nu-1}^{(1)}\left[k(\eta_e-\eta_{i})\right] - \frac{H_{\nu}^{(1)}\left[k(\eta_e-\eta_{i})\right]}{(\eta_e-\eta_{i})} \Bigg\rbrace \mu_{k}^{\mathrm{c}}(\eta_{i}) \nonumber \\
&&
-i\sqrt{\frac{\pi}{4}(\eta_e-\eta_{i})}H_{\nu}^{(1)}\left[k(\eta_e-\eta_{i})\right] \dot{\mu}_{k}^{\mathrm{c}}(\eta_{i}).
\end{eqnarray}
In the next section we will provide numerical values for the quantities that appear in these expressions, in the scenarios of phenomenological interest in LQC that we are considering in this work.

\section{Primordial power spectrum}

The analysis carried out so far provides all the ingredients for the computation of the primordial power spectrum of the scalar perturbations in hybrid LQC using (semi-)analytic formulas. In general, this quantity is defined as \cite{Langlois}
\begin{equation}\label{ppower}
\mathcal{P_{\mathcal{R}}}(k)=\frac{k^3}{2\pi^2}  \frac{|\mu^{\mathrm{ex}}_k (\eta_f)|^2} {z^2(\eta_f)}
\end{equation}
where $z=a^2\dot{\phi}/\dot{a}$, the functions $\mu^{\mathrm{ex}}_k$ form a complete set of exact positive-frequency solutions for Eq. \eqref{effMS}, and $\eta_f$ is any instant of time within or by the end of slow-roll inflation such that $k\ll \dot{a}(\eta_{f})/a(\eta_{f})$ for the observed modes. The arbitrariness allowed in the determination of this time is physically due to the fact that the evolution of the rescaled Mukhanov-Sasaki mode $v_{\vec{k}}/z$ freezes during slow-roll inflation after its corresponding wavelength has crossed the Hubble horizon \cite{mukhanov1}. 

In the previous section, Eqs. \eqref{musr}-\eqref{B} describe the evolution of the perturbations within slow-roll inflation, for the choice of an NO-AHD vacuum state in the epoch near the bounce in hybrid LQC. To compute the primordial power spectrum, as commented above, we need to evaluate those expressions at a time $\eta_f$ of the slow-roll inflationary epoch where the evolution of the relevant perturbation modes has frozen. Let us focus our attention on the window of modes $k\in [10^{-4},10^2]$. This window has been used in a number of studies in the literature \cite{NO,hybpred1,hybpred2,JCAPGB}. With sufficient flexibility to account for all the cases of interest, this window includes the curvature scales that are important in the effective LQC scenarios that we are studying. For this window, reasonable values of the conformal Hubble parameter $\dot{a}/a$ (i.e., $a$ times the standard Hubble parameter) at, respectively, $\eta_f$ and $\eta_e$ are, e.g., $10^8$ and $10^{20}$ \cite{hybpred2}, while numerically we find that $\eta_{e}-\eta_f\approx 10^{-7}$. 

Thus, the argument of the Hankel functions in our analytic formula \eqref{musr} for the positive-frequency solutions $\mu_k^{\mathrm{sr}}$ during slow-roll inflation is much smaller than one, at least for all scales $k$ in the considered window. We can therefore employ the limiting behavior of these functions \cite{Abram} to obtain
\begin{equation}
\left|\mu^{\mathrm{sr}}_k(\eta_f)\right|^2  \approx  \frac{1}{4\pi}(\eta_e-\eta_f)\left|\Gamma(\nu)\right|^2\left[\frac{k(\eta_e-\eta_f)}{2}\right]^{-2\nu}|A_k -B_k|^2,
\end{equation}
where $\Gamma$ is the gamma function. So, we can finally approximate the primordial power spectrum as
\begin{equation}\label{appower}
\mathcal{P_{\mathcal{R}}}(k)  \approx  C_{\nu} k^{3-2\nu}|A_k -B_k|^2,\qquad C_{\nu}  \approx \frac{1}{2\pi^2}\frac{\eta_e-\eta_f}{a(\eta_f)^2 \epsilon_W}\left|\Gamma(\nu)\right|^2\left(\frac{\eta_e-\eta_f}{2}\right)^{-2\nu},
\end{equation}
because during slow-roll inflation it is well known that $z^2 \approx  a^2 \epsilon_W/(4\pi)$ \cite{Baumann}. The only remaining quantities that we need in order to calculate the power spectrum are the time difference $\eta_e -\eta_i$ and the slow-roll parameters $\epsilon_W(\eta_f)$ and $\delta_W(\eta_f)$, which are equal to each other in the case of the quadratic potential. Our numerical simulations show that, with our choice of $\eta_f$ and $\eta_e$ above, $\eta_e -\eta_i  \approx  7700$ and $\epsilon_W(\eta_f)=\delta_W(\eta_f)  \approx  1.27\times 10^{-2}$.

Before passing to a quantitative analysis of our results, an important remark is in order. The (semi-)analytic function by which we have approximated the behavior of the effective mass $s$ for the Mukhanov-Sasaki perturbation modes in hybrid LQC is discontinuous and/or not differentiable at the instants $\eta=0$, $\eta=\eta_t$, and $\eta=\eta_i$. Such discontinuities prevent one from properly applying the NO-AHD criterion to select an optimally adapted vacuum state in the whole period of evolution, tailored to a smooth dynamics till the end of inflation. Moreover, in Ref. \cite{JCAPGB} it was argued that they are responsible for the appearance of spurious oscillations in the power spectrum, coming from the introduction of a dephasing between the constants of integration as the positive-frequency solutions adjust themselves to the sudden change in the mass at the discontinuities. A straightforward way to remove this somewhat artificial dephasing was also discussed in Ref. \cite{JCAPGB}. It consists in applying just a Bogoliubov transformation to the positive-frequency solutions during the slow-roll regime. This transformation is
\begin{equation}\label{Bog}
A_{k} \rightarrow \tilde{A}_{k} = |A_{k}|, \hspace{1cm} B_{k} \rightarrow \tilde{B}_{k} = |B_{k}|.
\end{equation}
Motivated by these arguments, in the remainder of this work we will employ the following expression for the primordial power spectrum, which eliminates in a neat way the spurious oscillations arising from our approximate (non-smooth) treatment of the cosmological dynamics:
\begin{equation}\label{NOpower}
\tilde{\mathcal{P}}_{\mathcal{R}}(k) = C_{\nu} k^{3-2\nu}\left(|A_k| -|B_k|\right)^2.
\end{equation}

\subsection{Quantitative results}

In the following, we evaluate Eq. \eqref{NOpower} for the primordial power spectrum with the choice of constants $A_k$ and $B_k$ that we have determined by adopting an NO-AHD proposal to select a vacuum in hybrid LQC. We use a quadratic potential of the form $W(\phi)=m^2\phi^2/2$, and focus our study on the specific values $m=1.2\times 10^{-6}$, $\phi_0=1.22$, $\eta_e-\eta_{i}=7700$, and $\epsilon_W=\delta_W=1.27\times 10^{-2}$ motivated in our discussion above. Moreover, to compute the first order correction in the potential to the effective mass during the relativistic epoch of kinetic dominance, we take the values of $V_0$ and $\mathcal{H}_{0}^{(F)}$ provided by Eqs. \eqref{matchV} and \eqref{matchH}.

\begin{figure}[t]
 \includegraphics[scale = 0.48]{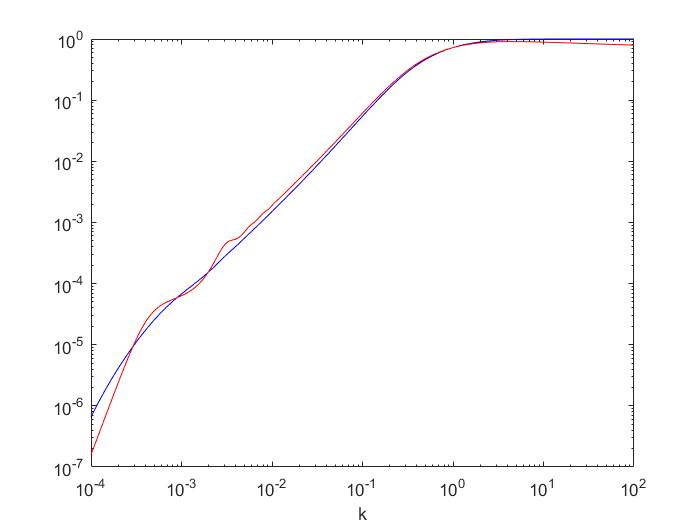}\includegraphics[scale = 0.48]{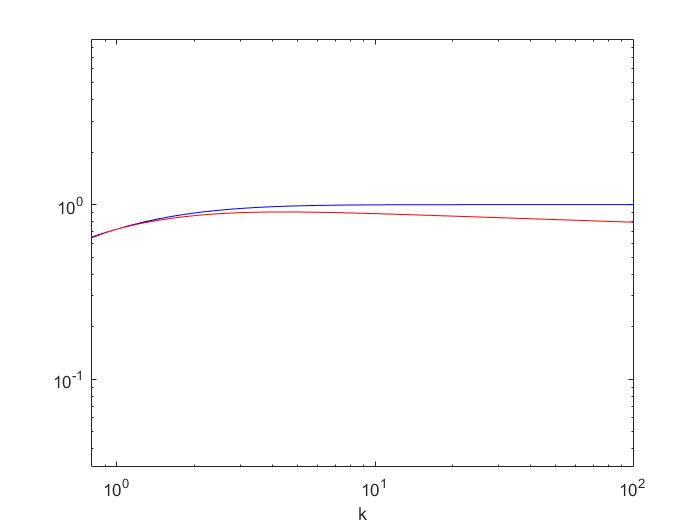}
 \caption{\textit{Left}: Comparison between the (normalized) primordial power spectrum $C_{\nu}^{-1}\tilde{\mathcal{P}}_{\mathcal{R}}$ obtained in this  work (red) and its analog without any contribution of the inflaton potential (blue). \textit{Right}: Zoom of the left panel in the region of large scales $k\in [1,100]$. In both panels, the axes are in a logarithmic scale. }
\label{fig2}
\end{figure}

The left panel of Fig. \ref{fig2} displays the resulting power spectrum in red. Superposed to it, we show in blue the result obtained after adopting the same criterion for the choice of an NO-AHD vacuum, but completely ignoring the effects from the inflaton potential both in the preinflationary epoch (including the transition period of constant mass) and during slow-roll inflation. The behavior of the two displayed spectra is very similar for scales $k$ in the interval $[10^{-2},k_{\mathrm{LQC}}]$, where $k_{\mathrm{LQC}} \approx  3$ is the curvature scale at the bounce (see e.g. the discussion in Ref. \cite{JCAPGB}). In this window, both spectra show power suppression to the infrared, which is approximately of exponential type. It has been long argued that such suppression may alleviate the anomalies observed in the CMB if one adheres to the standard cosmological paradigm, as long as the duration of inflation is such that the curvature scale $k_{\mathrm{LQC}}$ falls into the observational window. For scales $10^{-3}<k<10^{-2}$, on the other hand, the spectrum displays some features in the suppression of power when it includes the corrections discussed in this work. In fact, these features appear at a scale which is of the order of the square root of the constant mass of the second epoch of our approximation, and which is of the same order as the value of the conformal Hubble parameter at the onset of inflation \cite{JCAPGB}. Moreover, the power suppression for wavenumbers below $10^{-3}$ (although faster than for larger $k$ in both cases) is slightly more pronounced with the considered corrections than in the power spectrum computed without them. Finally, for scales greater than $k_{\mathrm{LQC}}$, we observe that both power spectra exhibit a (quasi) constant behavior that agrees with that of the standard Bunch-Davies state. Notice, however, that the consideration of slow-roll effects in this work introduces a red tilt, in contrast with the pure scale invariance that is obtained when treating the inflationary phase as an exactly de Sitter scenario. This red tilt is just a property of the quadratic potential studied here. We show it zoomed in the right panel of Fig. \ref{fig2}.

\begin{figure}[t]
	\includegraphics[scale = 0.48]{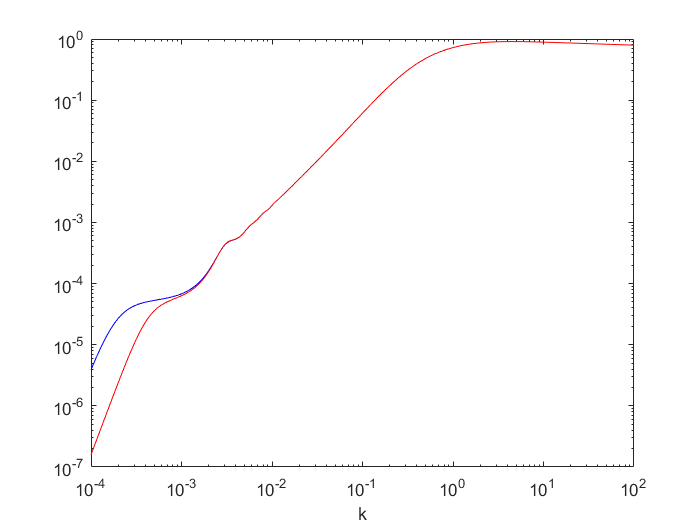}\includegraphics[scale = 0.48]{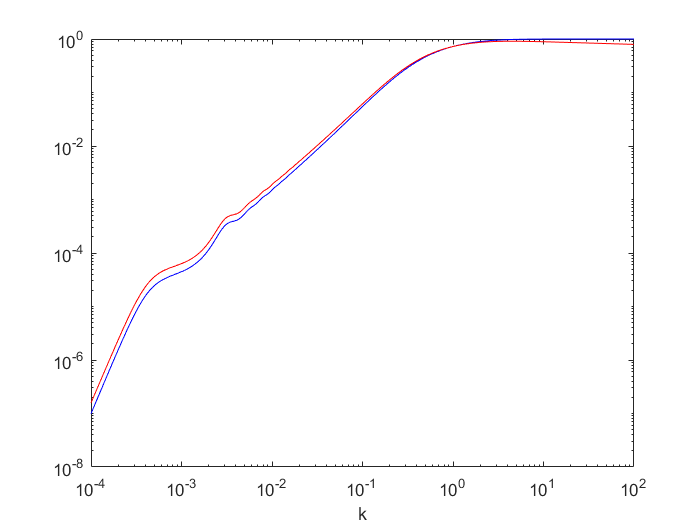}
	\caption{\textit{Left}: Normalized power spectrum $C_{\nu}^{-1}\tilde{\mathcal{P}}_{\mathcal{R}}$ obtained with all the modifications from the inflaton potential studied in this work (red) versus the result with $\mathfrak{F}_k^{(W)}=0$ (blue). \textit{Right}: Comparison between our normalized power spectrum computed with $\nu=(9/4 +9\epsilon_W -3\delta_W)^{1/2}$ for slow-roll inflation (red) and the spectrum with $\nu=3/2$ for de Sitter inflation (blue). In both panels, the axes are in a logarithmic scale. }
	\label{fig3}
\end{figure}

Given the commented differences between the primordial power spectrum computed with and without the corrections introduced in this work, it is interesting to understand the origin of each of these modifications. For this purpose, in the left and right panels of Fig. \ref{fig3} we respectively compare our results, on the one hand, with the spectrum obtained if we ignore the first order corrections $s^{(W)}$ to the effective mass in the kinetically dominated epoch (namely, if we set $\mathfrak{F}_k^{(W)}=0$ in Eq. \eqref{mukin}) and, on the other hand, with the spectrum reached if we neglect the slow-roll effects and set $\nu=3/2$ in our formulas, corresponding to an exact de Sitter inflation. The left panel clearly shows that the presence of $s^{(W)}$ is the main actor behind the slight increase in the rate of suppression for $k<10^{-3}$ found in Fig. \ref{fig2}. The right panel, on the other hand, indicates that slow-roll effects barely modify the spectrum in the window of scales $[10^{-4},k_{\mathrm{LQC}}]$. Remarkably, this window includes wavenumber scales between $10^{-3}$ and $10^{-2}$, for which Fig. \ref{fig2} shows a small bump when compared to the spectrum derived without any of the corrections from the inflaton potential introduced in this work. One can convince oneself that this feature arises from the transition epoch between kinetic dominance and inflation, which we modeled by introducing a constant value  $s_c=1.83\times 10^{-6}$ for the effective Mukhanov-Sasaki mass. Once this fact is realized, we inmediately confirm that the main consequence of the slow-roll effects is indeed the red tilt in the ultraviolet sector. Thus, we conclude that the modifications produced by the slow-roll regime in the shape of the primordial power spectrum can be incorporated in scenarios from LQC exactly as they are treated in the standard cosmological paradigm, i.e., by multiplying the spectrum obtained in de Sitter by a factor $k^{3-2\nu}$. 

\section{Conclusions}

We have carried out a semi-analytic study of the evolution of primordial fluctuations of scalar type in some physically interesting scenarios arising in hybrid LQC. As a main novelty with respect to previous works on this topic, we have introduced the most relevant modifications to the dynamics of the perturbations that are due to the presence of an inflaton potential. This has been done in three steps. First, we have explicitly computed the first order correction caused by the potential in the effective Mukhanov-Sasaki mass of the epoch after the LQC bounce, where the background geometry follows approximately the same dynamics as in GR and the energy density of the inflaton is dominated by its kinetic contribution. This correction arising from the potential grows as time passes. To incorporate it, we have particularized the general results of Ref. \cite{Rafael} to a classical regime with kinetic dominance, working out all computations in the specific case of a quadratic inflaton potential and completing all calculations explicitly. Eventually, the kinetically dominated epoch gives way to an inflationary phase of the Universe. Comparing the resulting approximation to the mass with its exact numerical counterpart, we have argued in favor of a transition period between the epoch of kinetic dominance and inflation, at times in the cosmological evolution when including the potential as a small correction is no longer valid. As a second step in our approach, this transition has been modelled just by setting the mass equal to a constant, with a precise value that is determined numerically. Finally, we have approximated the whole inflationary epoch by a slow-roll regime, computing the dynamics of the perturbations to first order in the slow-roll approximation.

Any complete study of the dynamics of the Mukhanov-Sasaki perturbations requires (in a way or another) a criterion to choose their initial conditions, choice that amounts to select their vacuum state. In this work, we have adhered to the NO-AHD proposal, which has a fundamental motivation in hybrid (Loop) Quantum Cosmology and is conceived to favor desirable nonoscillatory properties for the corresponding positive-frequency solutions. Taking advantage of previous experience in approximating the dynamics of the perturbations around the kinetically dominated bounce (when the LQC modifications to GR are significant) \cite{JCAPGB}, we have applied the NO-AHD proposal to select a vacuum state in that earliest epoch. This provides initial data for the perturbations during the relativistic epoch of kinetic dominance, epoch in which we have solved the dynamics of the perturbation modes at first order in the inflaton potential. The resulting solution fixes in turn the solution for each perturbation mode during the transition epoch of constant mass, and eventually also during the slow-roll regime. Evaluation of these mode solutions at late enough times within the inflationary period gives the primordial power spectrum associated with our choice of vacuum state. In order to study its properties, one should bear in mind that our approximations (to reach an analytic resolution of the perturbation equations) introduce spurious oscillations in the norm of the solutions, owing to the (necessary but artificial) appearance of discontinuities in the effective mass and/or its time derivatives. Following the same philosophy as in Ref. \cite{JCAPGB} with respect to such an artificial non-smooth behavior, we have removed these spurious oscillations in a neat way before quantitatively evaluating the spectrum for our LQC scenario.

The evaluation of the power spectrum obtained with our analysis shows that the effects of the inflaton potential do not qualitatively change the power suppression previously found \cite{AshtPRLLast,Ashtekarlast,JCAPGB} for $k$ smaller than the scale $k_{\mathrm{LQC}} \approx  3$ in Planck units, confirming that this suppression is genuine of LQC scenarios and it starts at a scale that is directly related to the universal value of the spacetime curvature at the bounce. However, the corrections introduced in this work quantitatively change how this suppression behaves in the very infrared end of the spectrum and become also relevant for $k>k_{\mathrm{LQC}}$. For very infrared scales, there are relevant modifications arising from the effects of the inflaton potential in the preinflationary epochs. On the one hand, the transition period of constant mass is manifest via a small bump in the power spectrum for values of $k$ between $10^{-3}$ and $10^{-2}$. On the other hand, the importance of the potential at the end of the period of kinetic dominance is reflected as an increase in the power supression for scales even futher in the infrared regime. For wavenumber scales $k$ greater than the curvature at the LQC bounce, which lie in the quasi scale invariant part of the spectrum, slow-roll effects produce a red tilt in the considered case of a quadratic potential, just as they do in the standard cosmological model.

We have conducted our analysis in the framework of the hybrid formalism of LQC, with an inflaton potential given by a mass term, and a choice of initial conditions selected by the criteria explained in Ref. \cite{JCAPGB}. We expect that a similar study can be carried out in the dressed metric formalism using as starting point the case of vanishing potential during kinetic dominance, recently developed in Ref. \cite{AMMVB}. A generalization to other potentials also seems to be at hand, especially to potentials that depend exponentially on the inflaton \cite{IMM}, given that most of the integrals that arise in our calculations involve this kind of functional dependence. On the other hand, work experience indicates that the results will be very sensitive to the choice of vacuum state \cite{hybpred2}, something that can significantly affect e.g. the scale of power suppression (for instance, see the cases analyzed in Refs. \cite{JCAPGB,MJRMonteC}).

Our investigations open a road to find an analytic parameterization of the primordial power spectrum in terms of the free parameters of our LQC scenarios, namely, the initial value at the bounce of the inflaton field and its potential. These, in turn, directly determine how long the inflationary period lasts, and therefore fix the relative location of $k_{\mathrm{LQC}}$ in the observable window of scales nowadays. It is therefore of great importance to include them in any statistical analysis on the predictions of LQC in what respects the Early Universe. The formulas obtained here clearly provide a promising starting point for such a parameterization. Moreover, the methodology developed in this work can be directly applied to also study the evolution of the tensor modes of the perturbations. This would provide a way to analyze the effects of the inflaton potential on the primordial power spectrum of gravitational waves in cosmological scenarios inspired by hybrid LQC. Finally, it would be interesting to perform a similar study on the evolution of the primordial fluctuations according to the dressed metric approach to LQC, in order to eventually establish whether it can be distinguished from the hybrid formalism using cosmological observations.

\acknowledgments

This work was partially supported by Project No. MICINN PID2020-118159GB-C41 from Spain, Grants NSF-PHY-1903799, NSF-PHY-2206557, and funds of the Hearne Institute for Theoretical Physics. The authors are grateful to A.Vicente-Becerril for helpful discussions and comments.

\end{document}